\documentclass[12pt, a4paper]{article}
\usepackage{bbm}
\usepackage{graphicx}
\begin{document}
\begin{center}
{\bf{\Large{The Coevolution of Individual Economic Characteristics
 and Socioeconomic Networks}}}
\end{center}

Dietrich Stauffer$^1$, Martin Hohnisch$^{2,3}$ and Sabine Pittnauer$^2$

\noindent
$^1$ Institute for Theoretical Physics, Cologne University, D-50923 K\"oln, 
Euroland 

\noindent
$^2$ Research Group Hildenbrand, Department of Economics, University of Bonn,
Lenn\'estr. 37, D-53113 Bonn, Germany

\noindent
$^3$ Department of Mathematics, University of Bielefeld, D-33501 Bielefeld, 
Germany 

\noindent
e-mail: stauffer@thp.uni-koeln.de; Martin.Hohnisch@wiwi.uni-bonn.de;

\noindent
Sabine.Pittnauer@wiwi.uni-bonn.de

\bigskip

Abstract: The opinion dynamics of economic agents is modeled with the link 
structure influenced by the resulting opinions: Links between people of 
nearly the same opinion are more stable than those between people of vastly
different opinions. A simple scaling law describes the number of surviving
final opinion as a function of the numbers of agents and of possible opinions.

\section{Introduction}
Local interaction structures, embodied in models of socioeconomic networks, have become increasingly
recognized in economics as an extension of global interaction mechanisms. 
In this literature, economic networks are usually taken as exogenous, say a square lattice or
a more complex graph.
In this note we make an attempt to make the structure of links between agents endogenous,
dependent on the degree of ``similarity'' between each pair of them.
The point of departure is a random graph structure modified by the assumption that links 
associated with a node, i.e. an economic agent, are not entirely random but influenced 
by the characteristics of other agents in the neighbourhood of that particular agent.

Simultaneously, we model the evolution of the characteristics themselves as a non-strategic social
adaptation process. For concreteness, we call the characteristics simply ``opinions''.
While this is not the standard terminology in economic literature, the reader will recognize
how the principle described extends to particular characteristics structures.

Our suggested process of opinion dynamics is based on previous work by \cite{Deffuant}.
Unlike in previous research, by the specification of endogenous links the network becomes 
dynamic, influencing the opinion dynamics and being influenced by it simultaneously.
 
Our investigation is based on computer simulation techniques using random numbers. Such an approach
 has a long history in economics
\cite{Stigler}.

\section{Model}
Our model uses simulation techniques known from the ``sociophysics'' literature
of opinion dynamics \cite{Axelrod} and Erd\"os-R\'enyi networks 
\cite{Bornholdt}. Each of $N$ agents (we used $N = 10^2, \, 10^3, \, 10^4$) can
have one of $Q$ opinions $(10 \le Q \le 10^4$). The opinion of agent $i$ is represented
by the variable $S_i$ taking values in a finite subset of $[0,1]$ consisting of
the numbers $n/Q$ with a natural number $n \leq Q$. 
At the outset, to each agent there is associated a reference group
generated by a repeated random selection of agents (repeated ten times in our simulations).
The  generated link structure is assumed to be one-sided, i.e. if $b$ is a reference person 
of $a$ then not necesserily vice versa. For example, they may 
represent relations between agents and their superiors.   

As in  the model of Deffuant et al \cite{Deffuant,Ben-Naim}, at every iteration each 
randomly selected agent $i$ discusses successively with the agents in its reference group.
In each instance, the two compare their opinions $S_i$ and $S_j$. 
If their opinion difference $|S_i - S_j|$ is larger than a fixed 
confidence interval $L$ (mostly between $Q/20$ and $Q/2$), they ignore each 
other's opinion; otherwise the two opinions move towards each other by an
amount  $|S_i - S_j|/\sqrt{10}$, rounded to the nearest integer value. (If their
opinions agree, nothing changes; if their opinions differ by only $\pm 1/Q$, one
of the two agents, randomly selected, adopts the opinion of the other.) This
discretization of opinions to $S = n/Q$ with natural numbers $n$ between 1 and $Q$, instead of 
continuous numbers $S$ between 0 and 1, is taken from \cite{Stauffer} to improve 
computational efficiency; for the same reason we did not use the alternative
model \cite{Hegselmann} where each agent looks at all $N$ agents instead of 
only ten of them. 

In addition, we allow for noise 
representing the random influence of the environment (good or bad economic 
news) \cite{HMO}, in addition to the opinion dynamics in the reference group. 
With probability 
1/2, each agent shifts its opinion randomly up or down by $\pm 1/Q$ (but
stays within the interval from 0 to 1). 

The coupling between the existence of a link between 
two agents and their opinions is to our knowledge the new aspect of our model: 
large differences of opinion destroy a link. Thus at
each iteration, before the above opinion dynamics starts, the reference group of 
each agent $i$ is reviewed.
The link to agent $j$ in the reference group is kept with a probability
$(p/Q)/|S_i - S_j|$ (if $S_i = S_j$ the link is kept with probability $1$), where $p=1/10, \, 1/2, 
\, 1$ was simulated ($p=1$ in all our figures). If a link is destroyed,
another bond is selected randomly; if the two opinions of this new bond are
far away from each other, this new bond will hardly survive the next iteration.

The simulations were continued either up to a fixed number $10^4$ of 
iterations or until a fixed point is reached. (Each agent is treated on average
once at each iteration; the number $t$ of iterations 
thus measures the time.) We define a fixed point as a situation when without
noise for ten consecutive iterations no opinion changed. With noise a fixed 
point is defined as a situation where due to the opinion dynamics, ignoring
the noise, during one iteration no opinion changed. The opinions at this fixed 
point are called final and are analyzed, with previous opinions ignored. 
Under conditions where noise prevents a fixed point to be reached within $10^6$ 
iterations, we make only $10^4$ iterations and average over all fluctuating 
opinion distributions in the second half of the simulations, $5,000 < t \le
10,000$.
 
Without noise we always found fixed points; with noise depending on parameters
we found fixed points, or we found opinions fluctuating about some stationary
distributions. In most cases we averaged over 1000 samples to get smoother 
statistics; the Fortran program of about 140 lines is available as deffuant19.f
from stauffer@thp.uni-koeln.de, as well as some figures mentioned but not shown
below. 

\section{Results}
\begin{figure}[hbt]
\begin{center}
\includegraphics[angle=-90,scale=0.45]{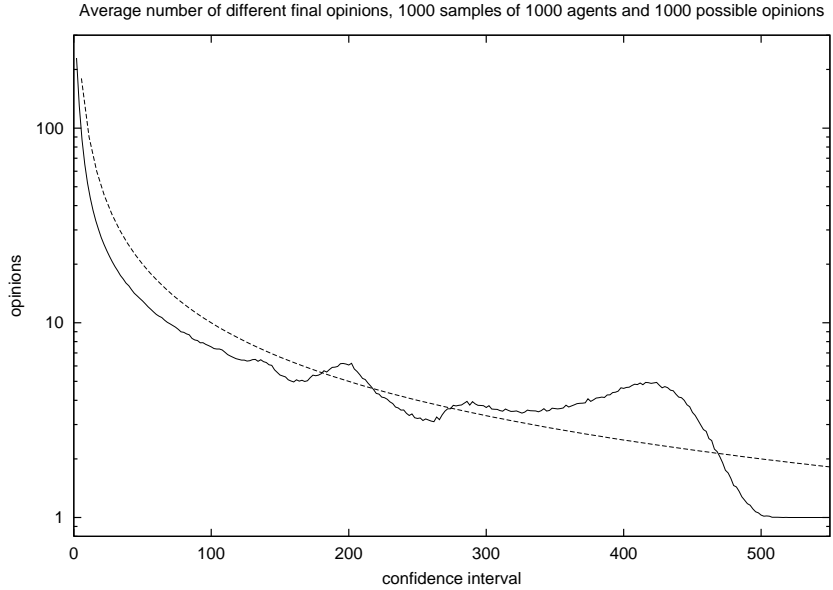}
\end{center}
\caption{
Semilogarithmic plot of the number of final opinions versus confidence interval
$L$; for $L > Q/2$ (here = 500) a complete consensus for only one centrist
opinion is seen.
}
\end{figure}

\begin{figure}[hbt]
\begin{center}
\includegraphics[angle=-90,scale=0.45]{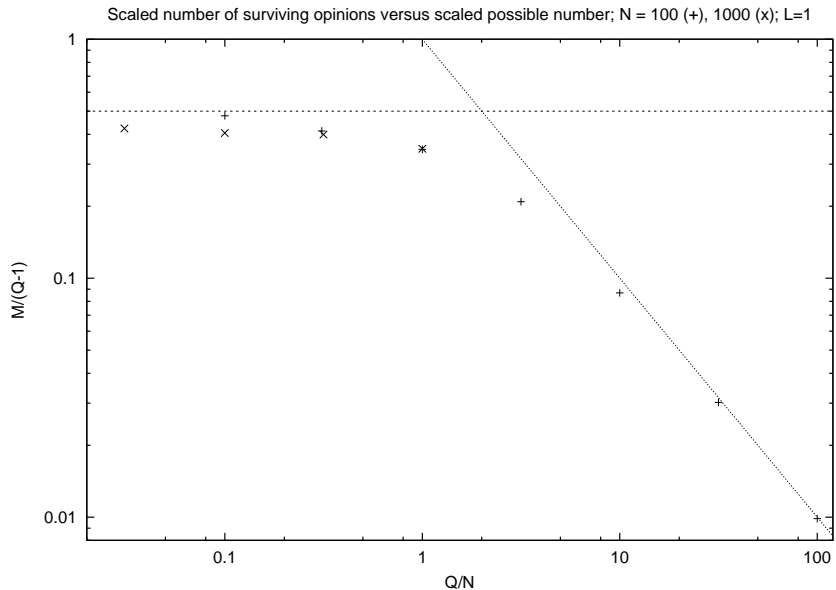}
\end{center}
\caption{
Double-logarithmic scaling plot. Different parameters $N = 100$ and 1000 lead
to the same curve.
}
\end{figure}

\begin{figure}[hbt]
\begin{center}
\includegraphics[angle=-90,scale=0.45]{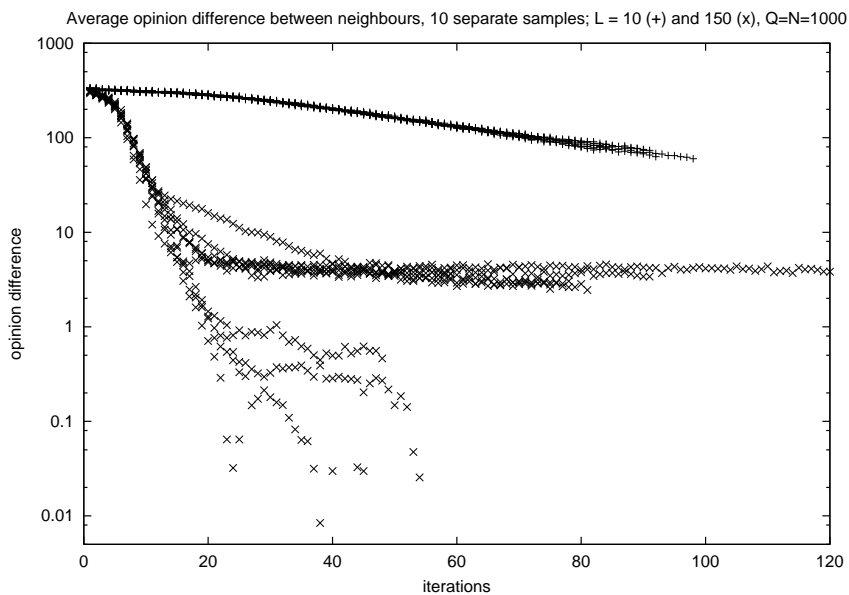}
\end{center}
\caption{
Semilogarithmic plot of the average opinion differences within  a link,
versus time. The upper data have a low confidence interval $L=10$, the lower 
data an intermediate $L=150$, for $Q=1000$. Ten different simulations are
shown separately (i.e. not averaged over). The opinions are multiplied here by $Q$ and thus vary from 1 to 1000.
}
\end{figure}

\begin{figure}[hbt]
\begin{center}
\includegraphics[angle=-90,scale=0.45]{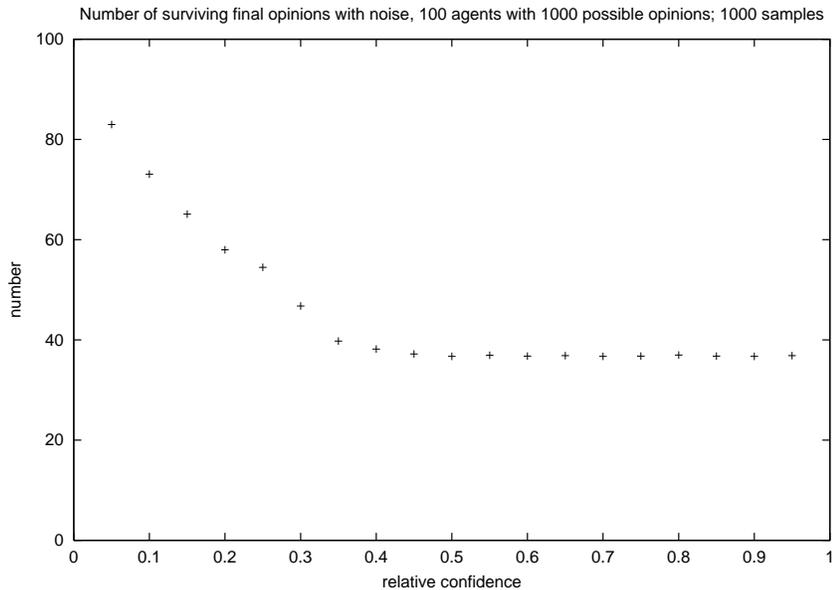}
\end{center}
\caption{
Linear plot of $M$ versus $L$ for strong noise; $N = 100, \, Q = 1000$.
}
\end{figure}

\begin{figure}[hbt]
\begin{center}
\includegraphics[angle=-90,scale=0.45]{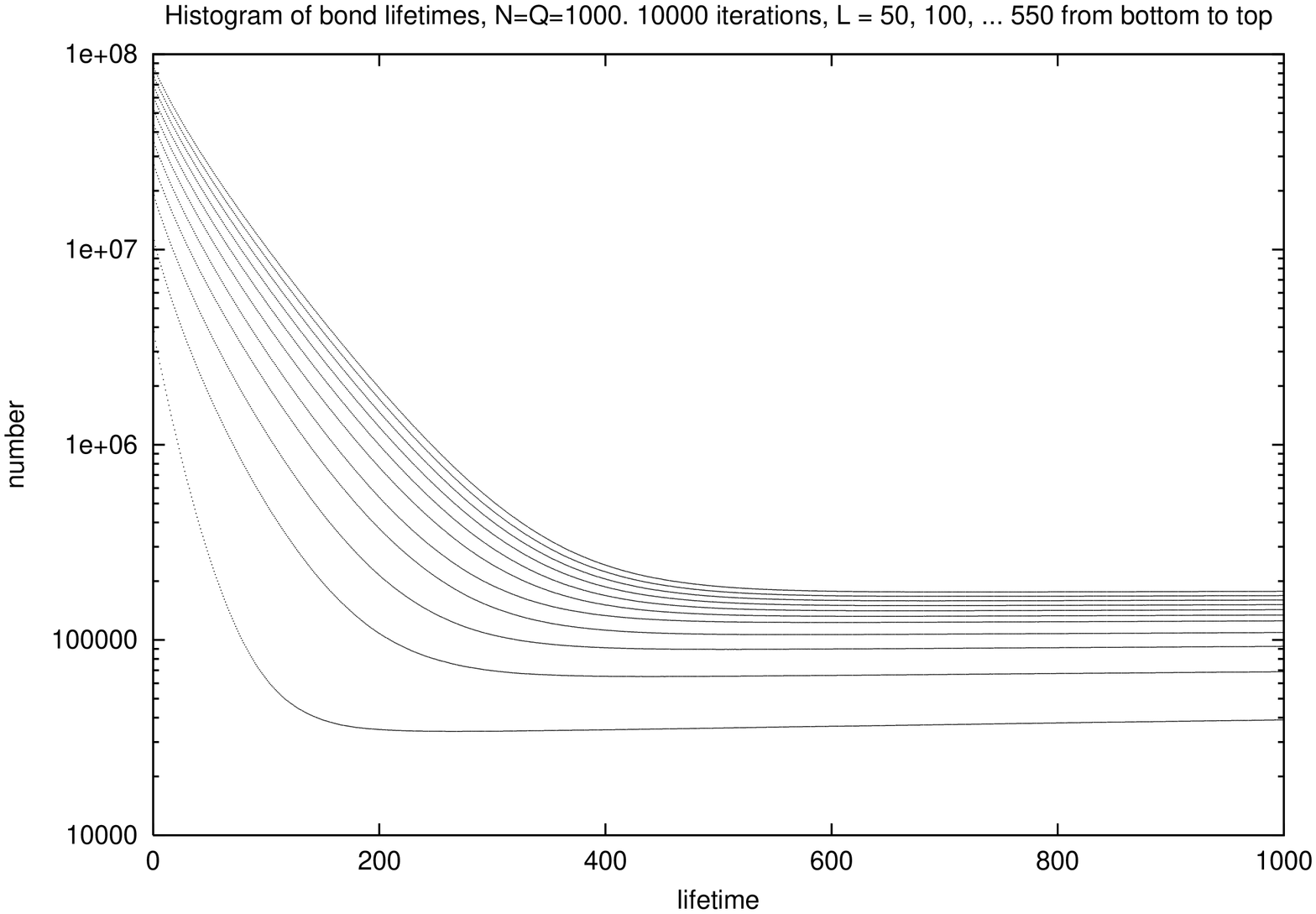}
\end{center}
\caption{
Semi-logarithmic histogram of lifetimes of links. Since each of the 
1000 agents in each of the 1000 samples contributes ten differences, the data
are much more smooth (``self-averaging'') than those in the earlier figures where
e.g. in Figure 1 the whole sample gave only one number $M$.
}
\end{figure}

\subsection{No noise}
Figure 1 shows that for large confidence intervals $L$ spanning more than half
of the possible opinion space a consensus is achieved: only one final 
opinion survives. For smaller $L$ more than one opinion can survive; we always
found a fixed point. The statistical fluctuations are barely visible, as was
shown by another simulation using the same parameters but different random
numbers. The deviations from the smooth curve $Q/L$ shown as a dashed line in 
Figure 1 are thus systematic; only for $2 \le L \le 50$ in the smooth left part 
the simulated results are proportional to $Q/L$. Similar data were obtained for 
different parameters $N, Q, p$, and also for the case without bonds where at 
each iteration each agent selects randomly one other agent for discussion
(not shown). 

In physics since 40 years many quantities were fitted on scaling laws. Thus a 
function $z=f(x,y)$ of two variables $x,y$ often can be written (for very small
or very large $x$ and $y$) as a scaled variable $z/x^a = F(y/x^b)$ given by
a function $F$ of only one scaled variable $y/x^b$, where $a,b$ are free
parameters often called critical exponents. In our case these exponents both 
are one, and the number $M$ of surviving opinions, scaled by $Q$, is a 
function of the scaled variable $N/Q$. Figure 2 shows this scaling function
in the form of $M/(Q-1)$ versus $Q/N$ for two drastically different system
sizes $N = 100$ and 1000 at $L = 1$: The two sets of data nicely overlap. For 
large $Q/N$ most opinions have no adherents, nearly $N$ different opinions have
one adherent each, and very few opinions have two or more adherents. Thus
only $M = N$ opinions survive in this limit, and the scaled variable $M/Q$
equals $1/(Q/N)$ as shown by the straight line with downward slope in Figure 2.
In the opposite limit of small $Q/N$, few agents initially share an opinion,
the small confidence interval $L=1$ allows many different final opinions 
separated by more than $L$, and thus the number $M$ of final opinions is 
proportional to $Q$: $M/Q$ = const in the left part of Figure 2, as indicated
by the horizontal line. The simplicity of this scaling law explains that it
is similar to the one found \cite{Stauffer} in a different model of fixed
links on a Barab\'asi-Albert network. (We divided $M$ by $Q-1$ instead
of by $Q$ since for Q = 2 a complete consensus $M = 1$ was found; for the
large $Q \ge 10$ used here the difference hardly matters.) For $L > 1$
a new parameter $L/Q$ would have to be used, and the scaling would have been
more complicated.
   
Figure 3 shows the dynamics until a fixed point is reached, at $Q = N = 1000$.
For large $L$ a complete consensus is reached as shown in Figure 1, and this
case therefore is less interesting now. For small confidence intervals $L$
we see in Figure 3 that the average difference of opinions with the ten
agents in the reference group decreases exponentially with time $t$. The ten simulated samples
give nearly the same results.  For intermediate $L = 150$ where on average 
about five opinions survive, we have drastic changes from sample to sample
even though they differ only by the random numbers used: Sometimes the  
opinion differences reach a constant plateau, and sometimes they nearly 
vanish. In spite of its simplicity the model thus indicates that one cannot
always predict that for two linked agents the opinions get close; one can
only predict that the opinions for two linked agents get closer than they were 
at the beginning. 

\subsection{With noise}
The constant noise disturbs the agreements which would have been found without
noise, and thus the number $M$ of final opinions in Figure 4 is much larger
than without noise: Instead of only one surviving opinion for large 
confidence intervals $L > Q/2$ we found on average 36.8 final opinions.
For small $L$ the number of surviving opinions is higher, as in Figure 1.
The parameters $N = 100, \, Q = 1000$ were chosen such that always a fixed 
point was found.

With $N = Q = 1000$ instead, no fixed point was found up to $t = 10^6$, and 
therefore we could look at the stationary distribution of the lifetimes for
each link. We see in Figure 5 that there are many links with a
lifetime of only one iteration; the number of observed lifetimes decays 
exponentially with increasing lifetimes, until for lifetimes of order
$10^2$ a plateau is reached, orders of magnitude below the maximum for 
unit lifetime.   

\section{Summary}
In this model of opinion dynamics,
for large enough confidence intervals $L > Q/2$ everybody finally
agrees with one centrist opinion, while initially the opinions were 
distributed randomly. In the case no such consensus is reached, the number
$M$ of surviving opinions obeys a simple scaling law, $M = Q \cdot F(N/Q)$, as
a function of the number $N$ of agents and the number $Q$ of possible opinions,
for large $N$ and $Q$. The consensus might correspond in reality to market
bubbles, like for information technology stocks before spring 2000, or for
tulips centuries ago. Our evolutionary process includes self-organisation of the
network of links between agents, depending on their opinions, and influencing
in turn their opinions. Improvements like inclusion of value judgments between
good and bad opinions, or influence of punctual events on the opinions,
are in preparation. 

MH gratefully acknowledges financial support from DFG grant TR 120/12-1, DS
that from COST P10 of the European Science Foundation.

\end{document}